# Low-Loss Integration of High-Density Polymer Waveguides with Silicon Photonics for Co-Packaged Optics

JEF VAN ASCH,[1,2,*] JEROEN MISSINNE,[1] JUNWEN HE,[2,3] ARNITA PODPOD,[2] GUY LEPAGE,[2] NEGIN GOLSHANI,[2] RAFAL MAGDZIAK,[2] HUSEYIN SAR,[2] HAKIM KOBBI,[2] SWETANSHU BIPUL,[2] DIETER BODE,[2] YOOJIN BAN,[2] FILIPPO FERRARO,[2] JORIS VAN CAMPENHOUT,[2] AND GEERT VAN STEENBERGE[1]

[1]*Centre for Microsystems Technology (CMST), imec and Ghent University, Technologiepark 126, 9052 Gent, Belgium*
[2]*Imec, Kapeldreef 75, 3001 Leuven, Belgium*
[3]*Currently at Huawei Technologies*
*\*jef.vanasch@imec.be*

**Abstract:** Co-Packaged Optics applications require scalable and high-yield optical interfacing solutions to silicon photonic chiplets, offering low-loss, broadband, and polarization-independent optical coupling while maintaining compatibility with widely used approaches for electrical redistribution. We present two heterogeneous integration techniques that enable high-density electrical and optical I/O connections, utilizing adiabatic coupling between on-chip silicon nitride (SiN) waveguides and package-level polymer optical waveguides. In the first approach, polymer waveguides are patterned using standard lithography directly on the surface of the photonic chip, ensuring compatibility with chip embedding as commonly employed in chip-first fanout wafer-level packaging. In the second approach, photonic chips are flip-chip bonded to the package substrate. Both techniques have been experimentally validated, achieving a coupling efficiency near 1 dB between SiN and polymer waveguides in O-band, for both TE and TM polarizations. SiN tapers were designed using the "Mono" method to optimize phase-matching conditions between the two waveguides, a critical requirement for integrating diverse optical components. These results demonstrate the potential of polymer waveguides in Co-Packaged Optics applications, achieving sub-2 dB chip-to-chip and chip-to-fiber coupling losses.

## 1. Introduction

The global demand for data transmission continues to grow exponentially, driven by the increasing requirements for high-performance computing (HPC) and artificial intelligence (AI) applications. This surge in data traffic presents significant challenges for datacenters and computational networks, which are under pressure to deliver higher bandwidths with lower power consumption and reduced latency. Up to now, datacenters have relied on conventional pluggable optics as primary means of interconnecting nodes. However, these traditional solutions are becoming increasingly inadequate, as they fail to meet the growing demands for bandwidth density. Consequently, conventional pluggable optical modules are becoming unsustainable for future datacenter applications that will experience even greater traffic volumes [1] [2].

To address these limitations, high-density Co-Packaged Optics (CPO) has emerged as an innovative solution. CPO facilitates the heterogeneous integration of optical and electronic components onto a single package substrate, substantially reducing the transmission distances associated with electrical channels on PCBs used for pluggable optics [3]. This integration leads to a more compact form factor, better energy efficiency and reduced cost, which is crucial for the scalability of next-generation systems. [4]. In this approach, conventional pluggable optics are replaced by optical engines, where optic-electric (OE) and electric-optic (EO) transitions are co-located with the central processor [5]. By integrating the optics closer to the central processor (e.g. Ethernet switch), the need for long electrical interconnections on the PCB is eliminated. An example configuration is depicted in Fig 1, where an Ethernet switch is surrounded by optical engines that each consists of an electronic integrated circuit (EIC) and photonic integrated circuit (PIC) sequence.

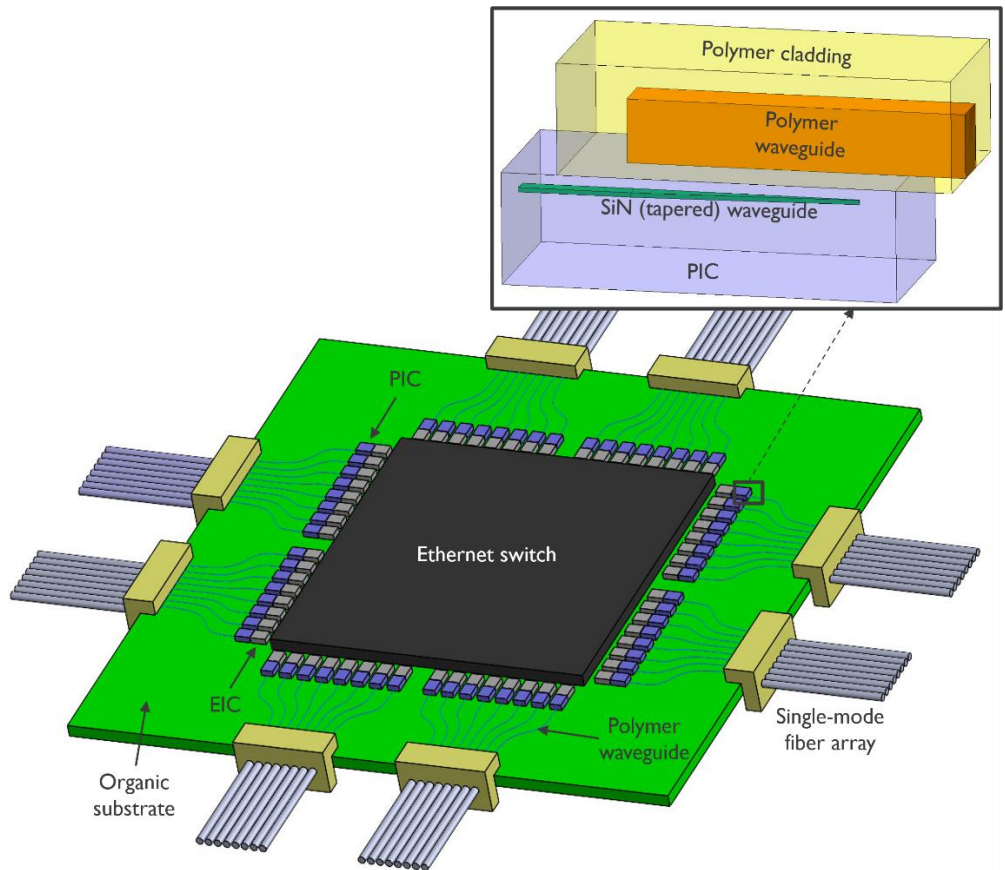

Fig 1. Visualization of the exemplary Co-Packaged Optics configuration for Ethernet networking: Optical engines consist of EIC-PIC sequence near the central processor. Polymer waveguides function as optical redistribution layer (ORDL) to establish optical connection between single-mode fiber arrays and PICs.

To integrate the optical engines, various packaging approaches are being developed that offer the required flexibility in their deployment for future high-density systems. A first method is to embed the PIC within an epoxy mold compound (EMC), alongside through mold vias (TMVs). The front and back electrical redistribution layers (ERDLs) allow to connect the PIC with the EIC and the package or PCB [6]. An alternative approach is to flip-chip bond the PIC on the optical interposer/substrate. Relying on micro solder bumps or copper pillars, strong bonding between both surfaces can be acquired. [7]. To further facilitate the optical component integration in either approach, fan-out wafer-level packaging (FOWLP) shows significant promise. Having shown to be effective for electrical interconnections, the development of optical FOWLP is now critical for scalable optical interfacing [3] [5]. Integrating an optical redistribution layer (ORDL) within FOWLP could streamline optical signal distribution across multiple chips on the same FOWLP. Techniques such as chip embedding or flip-chip bonding could enable the stacking of various optical elements on wafer-level, each serving a specific function, providing a scalable and cost-effective solution for future high-density optical systems [8] [9] [10].

Besides optical networking, CPO is also vital for the use case of optical compute interconnect (OCI) to support AI and machine learning (ML) applications [11]. To secure a high-bandwidth, low-latency, and low-power compute fabric, the currently used electrical interconnects are less likely to be sustainable in the future. Instead, wafer-level or package-level optical interconnects intend to combine the computation power of several XPU's along with its corresponding high bandwidth and low latency, as visualized in Fig 2. The energy efficiency requirement of the optical interconnects relies on low-loss optical coupling interfaces between the XPU and the interconnecting passive waveguide [12].

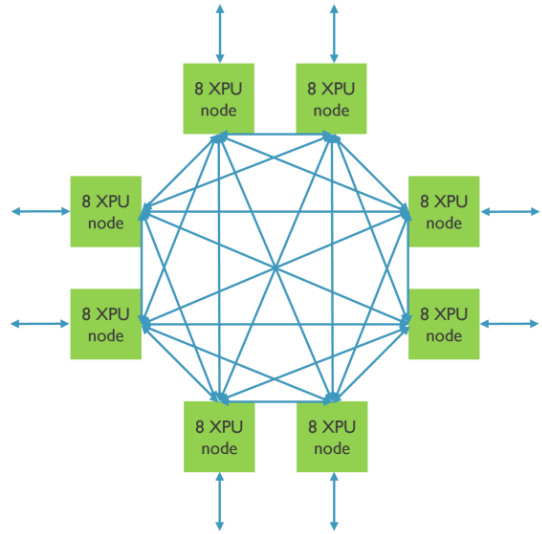

Fig 2. Graphical representation of an optical compute fabric, relying on multiple individual nodes interconnected to each other.

Polymer waveguides are particularly suited for ORDL due to their low propagation loss and compatibility with both wafer- and package-level implementations. For both CPO applications discussed above, polymer-based ORDL fan-out circuitry offers a scalable solution by integrating optical connections, thereby reducing the complexity and challenges associated with directly attaching an excessive number of optical fibers to the PIC. We highlight adiabatic coupling between on-chip SiN waveguides and single-mode (SM) polymer waveguides as an optimal approach for heterogeneous integration, ensuring high bandwidth capacity and energy efficiency. Analogous to electrical wire-bonding, adiabatic coupling is a surface-mediated optical interaction which enables 2.5D co-integration of electrical and optical components, interconnected by electrical redistribution layers. In this configuration, polymer waveguides not only serve as ORDL on the package substrate, but also act as mode converters, facilitating the coupling of optical signals to single-mode fiber (SMF) arrays.

So far, the co-integration of polymer waveguides on a PIC has relied on adiabatic coupling using Si tapered waveguides. The main challenge using SM Si waveguides is their difference in effective index of the fundamental TE and TM mode within the polymer waveguide. Consequently, small taper tips of less than 100 nm are required, which can lead to manufacturing issues. Since the adiabatic coupling efficiency heavily relies on the phase-matching of the confined modes in both Si and polymer waveguides, the polarization dependent loss (PDL) can vary for different taper lengths [13] [14]. To reduce the large refractive index contrast between Si and polymer waveguides, SiN is a valid material to be used as intermediate layer [15]. Going further, an inverse SiN taper is commonly employed to transition the large optical mode from flat-cleaved, industry-standard SMFs to the tightly confined mode in Si nano-waveguides, without sacrificing the simplicity of back-end-of-line (BEOL) integration [16]. Additionally, SiN waveguides possess a lower propagation loss and improved complementary optical properties compared to Si waveguides, making them more suitable for light routing. Simulations predict adiabatic coupling losses below 0.5 dB in O-band for 400 µm long tapers in case both SiN and polymer waveguides are tapered [17]. As a first experimental result, sub-1 dB adiabatic coupling loss was achieved between a 500 µm long SiN taper and rib-type polymer waveguide defined by contact lithography [18].

In the subsequent sections, we provide a detailed discussion of the proposed integration techniques, specifically lithography and flip-chip bonding. These methods facilitate face-up embedding and flip-chip soldering of the PIC onto the package substrate. To the best of our knowledge, this is the first experimental demonstration of adiabatic coupling between strip-type SiN tapers and the polymer-based ORDL in the O-band spectrum, utilizing the same SiN taper design for both integration techniques.

## 2. SiN-to-ORDL coupling concept

To allow for the most efficient adiabatic coupling, the confined optical mode needs to transfer from one waveguide to another with minimal conversion to higher-order or radiation modes. Our novel design of the SiN taper shape has been carefully optimized based on the "Mono" method [19]: high-efficient adiabatic coupling is acquired by monitoring the phase-matching condition between both waveguides. For both integration techniques, Fig 3(a)-(b) displays the change in effective index of the confined TE/TM modes in isolated SiN and isolated ORDL as a function of the change in SiN waveguide width. When both waveguides are included within the same simulation, the TE and TM supermodes jointly converge to the isolated ORDL's effective index with decreasing SiN width. This major advantage allows for the same SiN taper design on the PIC to support the optical coupling of light for both polarizations, as shown in Fig 3(c)-(d). Since the exact location of the phase-matching condition is almost identical for both integration techniques, the same taper design is used as well as slightly modified variations. The latter is incorporated as part of a design of experiments approach to estimate performance variations, accounting for potential minor perturbations in the SiN taper profile during the chip manufacturing process. Simulation results, preliminarily described in [20] prove that the highest coupling efficiencies are obtained by exploiting the phase-matching condition between both (tapered) waveguides.

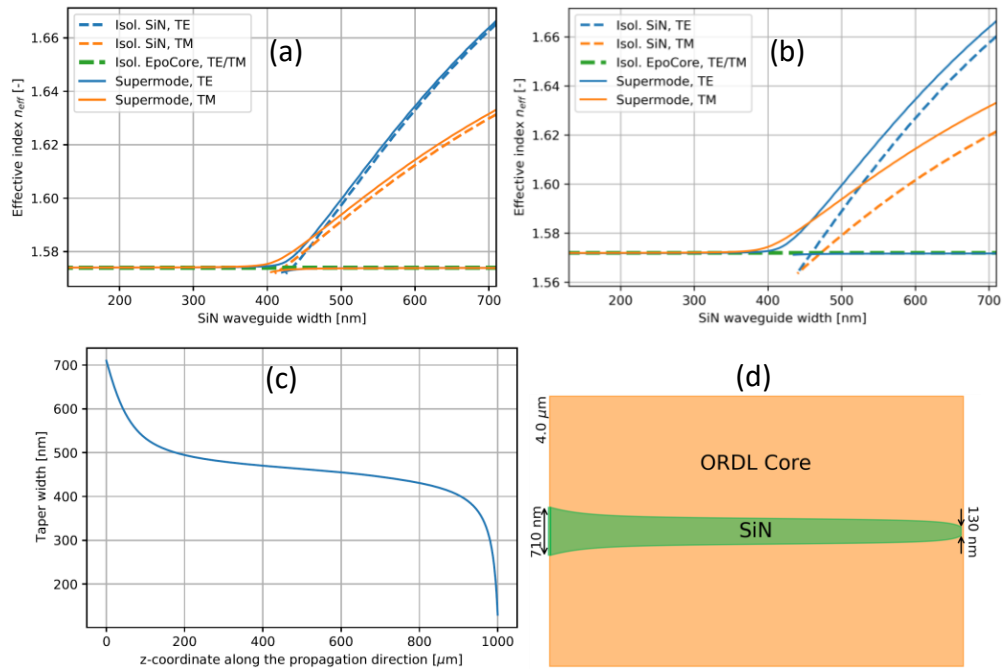

Fig 3. Lumerical Mode FDE simulation results, showing the effective index of the TE and TM guided modes inside the isolated SiN (RI = 2.0) and EpoCore (RI = 1.579) waveguides as well as the supermodes for the actual use case, as a function of varying SiN waveguide width for (a) lithography approach and (b) flip-chip bonding approach. (c) SiN taper width as function of z-coordinate along the propagation direction with a total length of 1 mm. (d) Bottom view of the SiN taper together with fixed ORDL core dimensions.

As a first indication, the Lumerical Mode eigenmode expansion (EME) simulations in Fig 4 display the optical coupling efficiency for both methods at discrete wavelengths in the O-band. The target value of 1 dB adiabatic loss between tapered SiN and the polymer ORDL at $\lambda = 1310$ nm can be reached for both TE and TM in case the taper length is at least 750 µm. To maximize optical coupling efficiency between the PIC and ORDL across a broad wavelength range, we have selected a minimal taper length of 1000 µm in the PIC test plan. Note as well in Fig 4 that the polarization dependent loss (PDL) is less than 1 dB for taper lengths larger than 1000 µm, for both the lithography deposition as flip-chip bond integration. Based on these promising simulation results, we fabricated the test samples as described in next section.

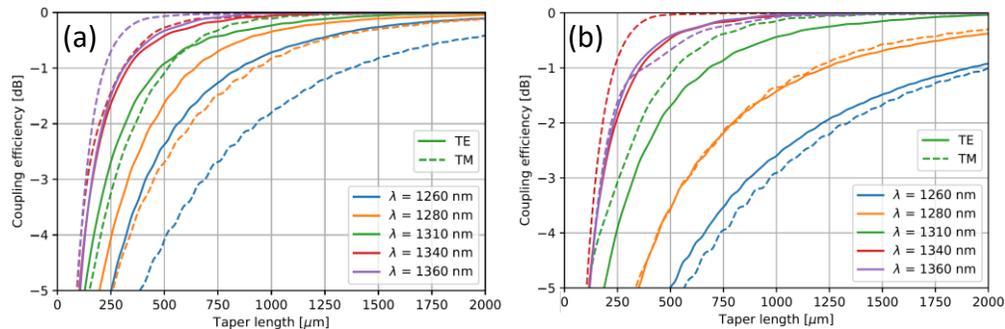

Fig 4. Lumerical Mode EME simulations show the optical coupling efficiency as a function of taper length, for 5 discrete wavelength values in the O-band: (a) Lithography approach. (b) Flip-chip bond approach.

## 3. Materials and methods

The silicon photonics chip consists of SiN waveguides, formed by Plasma-Enhanced Chemical Vapour Deposition (PECVD). The waveguides are 400 nm thick and have a nominal width of 710 nm, optimized for single-mode confinement in the spectral O-band. To facilitate optical coupling between the PIC and ORDL, the optimized adiabatic tapers integrated into the PIC feature a 130 nm taper tip while maintaining a uniform height of 400 nm along the entire taper length. The SiN layer is surrounded by a stack of multiple oxide layers, ending with a top oxide thickness of 25 nm, as shown in Fig 5(a)-(b). This is carefully chosen to eventually allow optical proximity between SiN and ORDL waveguides, while maintaining a planarized, protective layer on top of the SiN layer.

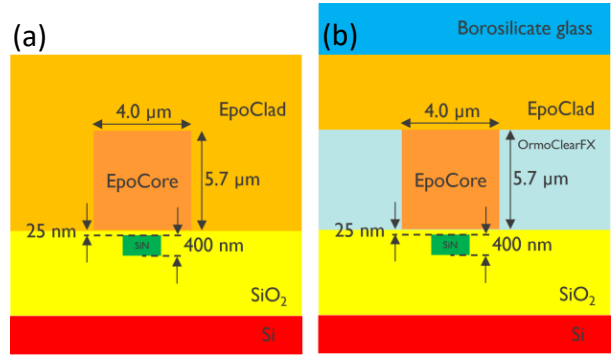

Fig 5. Designed cross-section of the various samples, highlighting all used materials. (a) Lithography approach. (b) Flip-chip bonding approach. [Dimensions are not to scale]

As a material for ORDL, commercially available EpoCore 5 (with refractive index (RI) = 1.579 at $\lambda$ = 1310 nm) is used for the core and EpoClad 20 (with RI = 1.571 at $\lambda$ = 1310 nm) as surrounding cladding [21]. In the following paragraphs we will discuss in more detail the 2 integration techniques used, for which the eventual cross-sections are shown in Fig 5. Considering the asymmetric cladding around the ORDL core, the EpoCore waveguides are designed to be 4.0 µm wide and 5.7 µm thick. These dimensions are selected to satisfy the SM condition and achieve a symmetrical mode profile with a mode-field diameter (MFD) of 6 µm.

### 3.1 ORDL deposition on embedded PICs

A first method to deposit ORDL waveguides on Si PICs, is by using lithography. This is mostly suitable for applications where the PIC is face-up integrated in a package. Prior to spin-coating the EpoCore 5 material at 2450 revolutions per minute (rpm), the PIC's top surface is activated by oxygen plasma treatment for adhesion improvement. After edge bead removal of the EpoCore 5 layer, a prebaking step is performed ramping up the temperatures from 50 °C to 90 °C for a total annealing time of 8 minutes. The lithographic deposition is done by a dedicated lithography mask and the EVG 620 mask alignment system. A long-pass filter is used to minimize T-topping effects associated with shorter wavelengths during the UV-exposure at a dose of 200 mJ/cm². A post exposure bake step is performed by analogously ramping up the temperatures again for 8 minutes, followed by a relaxation time of 30 minutes. The development is done using mr-Dev 600 [22] and a two-fold submersion in isopropyl alcohol (IPA). Finally, a hard bake step is performed at 120 °C for 1.5 hours to fully solidify the EpoCore waveguides.

Once the EpoCore waveguides are ready, the top surface is activated again by oxygen plasma treatment. EpoClad 20 is then deposited by spin-coating at 5000 rpm, covering both the top oxide layer and EpoCore waveguide pattern, acting as top and side cladding. The target EpoClad thickness is 10 µm, to sufficiently confine the modes in de ORDL core/cladding structure. The samples are heated up from 50 °C to 120 °C for 20 minutes in total. Then, UV-exposure is performed with an exposure dose of 600 mJ/cm², followed by the post exposure bake for 5 minutes at 90 °C. Finally, when the hard bake step is done at 120 °C for 1.5 hours, the samples are diced to access the EpoCore waveguides at both edges of the chip for loss assessment purpose. As an overview, Fig 6(a) schematically shows the result after performing all processing steps. The Epocore waveguide pitch is 50 µm, to fully support the FOWLP concept, enabling high waveguide density.

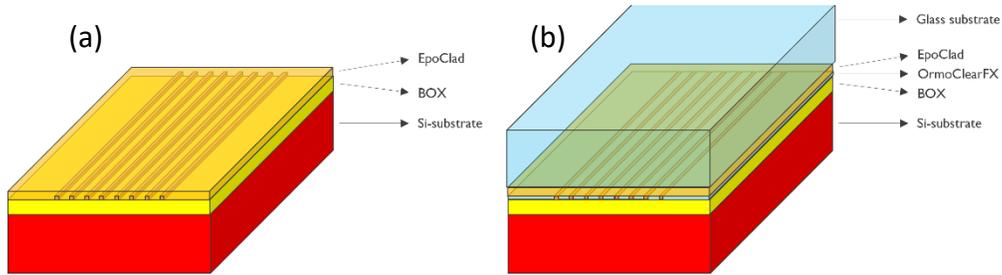

Fig 6. Illustration of the fabricated sample types, highlighting all used materials. (a) Lithography approach. (b) Flip-chip bonding approach. [Dimensions are not to scale]

### 3.2 Flip-chip bonding PICs on substrate

A second technique involves first depositing the EpoCore/EpoClad stack onto a separate substrate, which is then flipped and bonded to the PIC. While the PIC is typically bonded to the substrate, this reversed process is employed for practical advantages. This can also serve as an alternative coupling method in case a PIC is flip-chip integrated on a package substrate. In contrast to the lithography approach, EpoClad is first spin-coated onto a borosilicate glass plate, followed by baking, UV-exposure, and a second baking step using the same process parameters. Similarly, the EpoCore waveguide pattern is then established using the lithography procedure. The major difference with the lithography technique is that there is no side cladding yet at this stage, since the EpoCore waveguides are sticking out above the (quasi) planarized EpoClad layer.

The bonding between the top sample (glass plate with EpoClad/EpoCore) and the bottom sample (PIC) is done with the EVG 620 mask alignment system. First, the top glass sample is temporarily attached to a transparent glass mask by wax. The combined structure is then lifted by the EVG aligner to compensate for any tilt. Next, the PIC is positioned at the bottom substrate holder and 4 droplets (with diameter around 2 mm) of UV-curable OrmoClearFX (RI = 1.537 at λ = 1310 nm) are applied to the edges of the top surface. The 2 samples are brought into contact with each other to level out any additional tilt and then separated by 100 µm to perform the alignment. The passive alignment procedure relies on alignment markers that are present on both the SiN layer for the bottom sample and EpoCore layer for the top sample. Once the alignment between both samples is set correctly, they are brought into contact. The 4 droplets, smeared out meanwhile, act as glue once the UV-exposure of 1000 mJ/cm² by the EVG aligner is done through the glass substrate and mask. Afterwards, the bonded samples need to be released from the transparent glass mask by heating up the wax. In the current situation, the EpoCore pattern is confined by EpoClad at the top side, the PIC at the bottom side and air at either horizontal side. Therefore, additional OrmoClearFX is applied at one edge of the bonded sample to capillary fill the air voids between 2 adjacent EpoCore waveguides, demonstrated in Fig 5(b) and Fig 6(b). OrmoClearFX is a suitable material for this, as it does not contain any solvent and is viscous enough for capillary filling. The last step in the procedure is a final UV-exposure of 1050 mJ/cm², to fully cure the OrmoClearFX. The exposure is done inside a closed compartment filled with nitrogen gas, to avoid contact with air, which is beneficial for curing. When all steps are performed, the samples are diced to obtain access to the EpoCore waveguide facets. Both the glass and PIC are diced at the same position purely for testing purposes, while in the real applications the glass substrate (containing the EpoCore waveguides) can be kept larger.

### 3.3 Optical measurement setup

The measurement setup consists of two alignment stages with 6 degrees of freedom, allowing the fine alignment of the input and output fiber (standard SMF-28 fiber) with respect to the sample. First, the fiber-to-fiber transmitted power is measured and set as a reference, cancelling out all loss contributions not related to the samples. The wide-spectrum light source first passes by an O-band circulator, then through a polarization controller to ensure TE/TM polarization to finally reach the input fiber tip. Light is coupled to the desired ORDL waveguides, located at the edge of the sample, and propagates further into the sample. Depending on whether a functional test site (ORDL-to-SiN-to-ORDL as shown in Fig 7) or reference ORDL waveguide is being measured, a lateral offset of 50 µm between input and output facet is implemented in the SiN pattern on the PIC to cancel out any light tunneling through the sample. This means that 2 adiabatic transitions have been passed when testing the functional test sites. Once the light is coupled back to the output fiber, an optical spectrum analyzer (OSA) examines the wavelength-dependent output intensity.

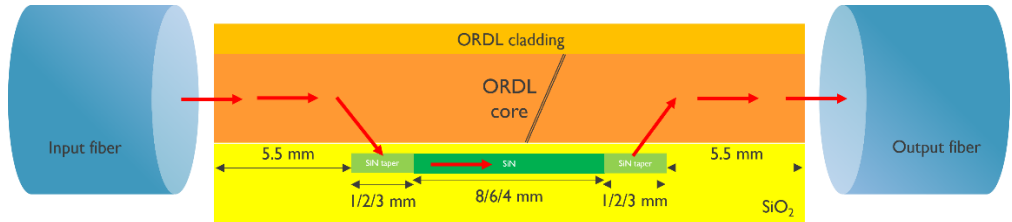

Fig 7. Schematic side view of a functional test site within a lithography-based sample. The red arrows depict the light travelling from input fiber, through the sample, and towards output fiber. [Dimensions are not to scale]

## 4. Experimental results

### 4.1 Lithographic approach

In total, six identical samples were produced for optical characterization, of which five were used for measuring the functional test sites containing adiabatic tapers and one was used for examining the ORDL propagation loss by cut-back analysis. The functional test sites (including "Mono" tapers) on the PIC were typically separated by 50 µm and each contained a slightly altered adiabatic taper design. All five functional samples had been diced at the same position, resulting in 21 mm long samples along the direction of the SiN and EpoCore waveguides. The microscope image of Fig 8(a) shows a cross-section, including all the present materials for the lithography deposition method. Fig 8(b) shows a SEM micrograph of a lithography sample without any top cladding. The ORDL core was deposited in the middle of the tiling-free region, on top of the SiN tapered waveguide (hence not visible on the micrograph).

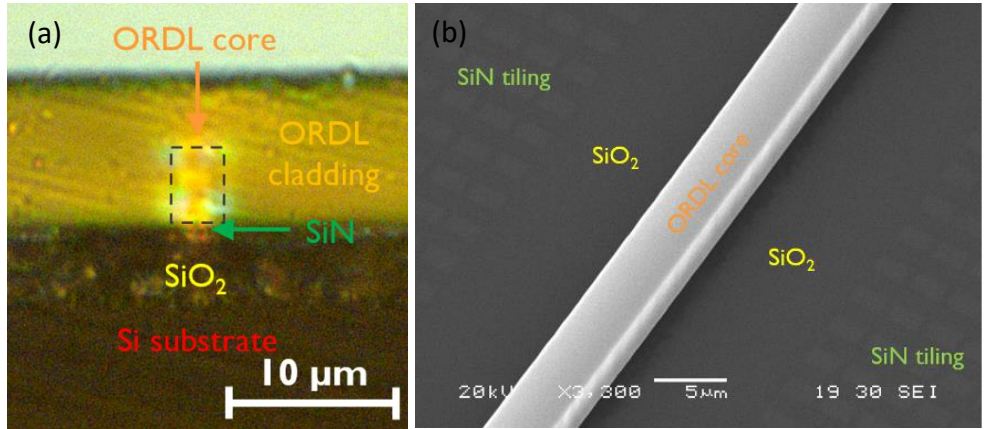

Fig 8. (a) Microscope image of a cross-section for one of the lithography samples. (b) SEM micrograph of an ORDL waveguide on a lithography sample before the top EpoClad layer is deposited.

The cut-back analysis was performed by consecutively shortening the sample by dicing perpendicular to the ORDL waveguides. For each configuration length, the ORDL reference waveguides (i.e. without any coupling to SiN) were optically characterized. Fig 9(a) shows the total coupling efficiency for each sample length, for both TE and TM polarizations at the averaged central wavelength region $\lambda = 1310$ nm $\pm 5$ nm to level out high-frequent changes. It can be observed that the ORDL propagation losses are 0.43 dB/cm $\pm$ 0.15 dB/cm for TE and 0.48 dB/cm $\pm$ 0.30 dB/cm for TM. Additionally, the butt-coupling loss per facet is 1.05 dB $\pm$ 0.18 dB for TE and 0.92 dB $\pm$ 0.36 dB for TM. These results match with the previously reported values in [23]. When performing the analysis over the entire O-band, the behaviour is shown in Fig 9(b). Additionally, by analyzing functional test sites with different SiN lengths, the SiN propagation losses are displayed as well in Fig 9(b) together with the 1$\sigma$ variation over the five considered samples, which are in line with previous results [24].

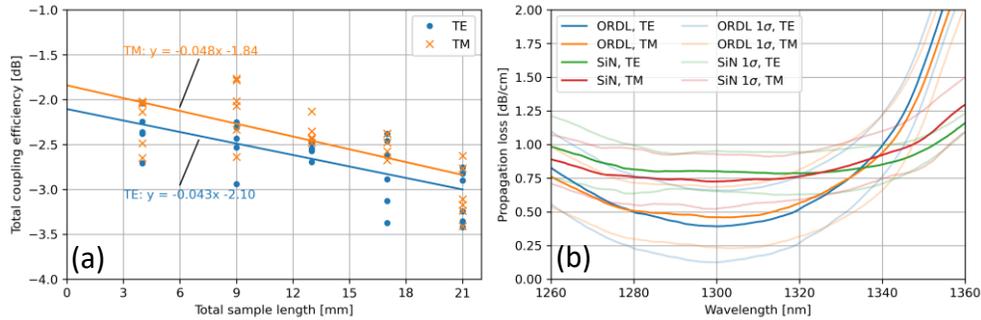

Fig 9. (a) Cut-back analysis to extract the ORDL propagation loss and butt-coupling loss, shown at $\lambda = 1310$ nm +/- 5 nm. (b) Propagation losses for ORDL and SiN waveguides, the latter including the 1$\sigma$ variation over five samples.

As final visualization of the loss breakdown analysis, Fig 10 shows the total insertion loss and each contribution, averaged over the five samples. The total insertion loss equals two times butt coupling (fiber-to-ORDL), ORDL propagation over a length of 11 mm, SiN propagation over a length of 8 mm and two times adiabatic coupling (SiN-to-ORDL) for 1 mm long tapers. At $\lambda = 1310$ nm, the adiabatic coupling efficiency equals -0.56 dB $\pm$ 0.37 dB for TE polarization and -1.04 dB $\pm$ 0.42 dB for TM polarization.

These results demonstrate the feasibility of achieving sub-2 dB chip-to-chip coupling loss, which includes two adiabatic transitions and assumes an ORDL propagation distance of less than 1 cm. Similarly, we calculate a sub-2 dB chip-to-fiber coupling loss, factoring in one adiabatic transition, ORDL propagation loss, and a sub-1 dB ORDL-to-fiber butt coupling loss. Important to note is that the sub-2 dB value does not rely on dedicated fibers for directly interfacing edge couplers, but only regular SM fiber. Hence, the eventual chip-to-fiber coupling loss can be optimized even further. An additional improvement could be achieved by tapering the ORDL as well, rather than only the SiN. These findings are crucial for advancing CPO, as they highlight the potential for efficient optical connections.

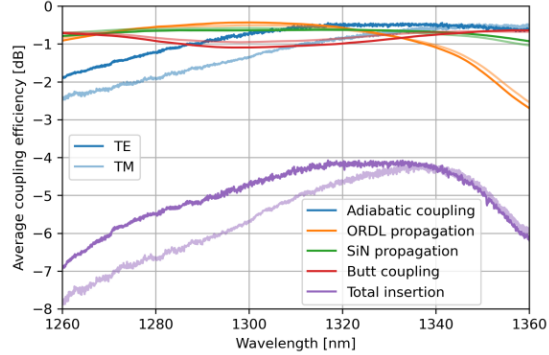

Fig 10. Average coupling efficiency over the five samples for each of the loss contributions.

### 4.2 Flip-chip bonding approach

Similar to the lithography samples, a batch of functional samples was fabricated as discussed in section 3.2 and further analyzed thoroughly by optical characterization. In addition, extra functional samples were produced with the same material stack but with dimensions suitable for inspection via the (destructive) Focused Ion Beam – Scanning Electron Microscopy (FIB-SEM) analysis. Besides a microscope cross-section in Fig 11(a), the SEM micrograph is depicted in Fig 11(b).

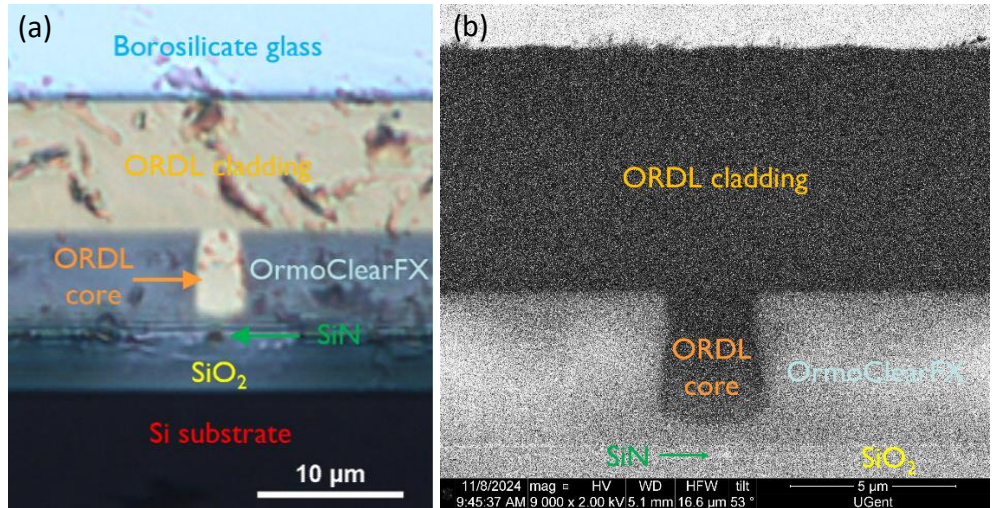

Fig 11. Cross-section for one of the flip-chip bonding samples. (a) Microscope image. (b) SEM micrograph.

Fig 12 presents the optical performance of the best-performing borosilicate sample, highlighting the total insertion loss. Although the data exhibits a noisier pattern compared to Fig 10, the overall coupling efficiency remains comparable to the lithography approach, with a maximum additional loss of 1 dB at $\lambda = 1310$ nm (TE/TM coupling efficiencies are -4.3 dB/-

5.0 dB for the lithography approach versus -5.2 dB/-4.9 dB for the flip-chip bonding approach). These results confirm that the flip-chip bond integration method is a viable alternative for SiN-to-ORDL coupling.

An attempt was made to conduct a full loss breakdown analysis using the cut-back method, but it yielded inaccurate results. Therefore, based on the findings of Fig 10 and Fig 12, it is more reliable to conclude that the adiabatic coupling efficiencies for flip-chip bond integration are comparable to those of the lithography approach, with an estimated additional loss of 1 dB.

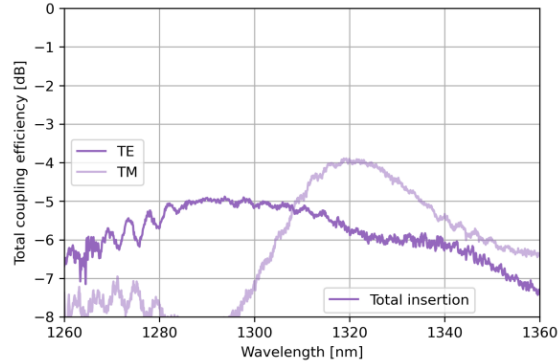

Fig 12. Total coupling efficiency of the well-performing flip-chip bonding sample.

## 5. Discussion

Adiabatic coupling between PIC and ORDL has been shown to be very effective, with coupling losses close to 1 dB for a taper length equal to 1 mm at $\lambda = 1310$ nm. Fig 13(a) shows more in detail the average coupling efficiency with the $\pm 1\sigma$ variation along the five lithography samples. The maximum coupling efficiency is achieved near $\lambda = 1320$ nm (1350 nm) for TE (TM) polarization. This red shift can be explained by SiN taper fabrication perturbations as well as altered EpoCore waveguide dimensions, compared to the original design. For the entire O-band, the polarization-dependent loss is lower than 0.8 dB, supporting both polarizations as information carrier.

Analyzing the tapers of 2 mm and 3 mm long in Fig 13(b), the efficiency values continue to improve with increasing taper length as expected. For both lengths, the efficiency ranges between -0.3 and -1.2 dB for both polarizations.

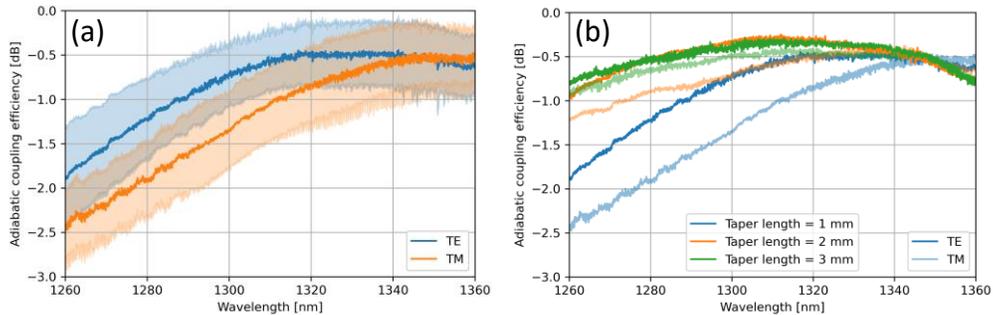

Fig 13. (a) Average adiabatic coupling efficiency for the lithography approach (taper length = 1 mm), including the $1\sigma$ variation. (b) Evolution of average adiabatic coupling efficiency as function of taper length.

Our results prove the versatility of adiabatic coupling in combination of different integration techniques, which can significantly impact future packaging architectures. However, one must be cautious when implementing these experimental implementations into application-driven setups for a few reasons.

Regarding the flip-chip experiments, a total amount of 16 samples were fabricated in three different production batches. Out of these 16, only two samples have shown acceptable results for the functional test sites containing the "Mono" tapers. Some samples performed well only at certain locations on the chip, whereas other failed to couple any light at all. This was mainly due to the poor bonding quality caused by warpage issues, resulting in a large vertical distance between both SiN and EpoCore waveguides.

Since 2 separate substrates are used, warpage plays a crucial role in the eventual bonding accuracy. Both the bottom sample (PIC) and top sample (glass) have a typical warpage of less than 3 µm over the entire length along the waveguides (2-3 cm) just before bringing them into contact with each other. The glass substrate's warpage depicts a convex pattern, due to a higher coefficient of thermal expansion (CTE) value for EpoClad (~87 ppm/K) [25] compared to borosilicate glass (~4 ppm/K) [26]. The PIC warpage shows a more irregular pattern compared to the glass. Consequently, this is the main challenge to bring both the SiN (tapered) waveguide and ORDL in close optical proximity. Therefore, the flip-chip bonding process in our tests is more prone to fabrication failures compared to lithography integration, resulting in poor reproducibility in the results. Lumerical Mode EME simulations show that the coupling efficiency is reduced by 10 dB in case the vertical SiN-to-EpoCore distance has an additional offset of 1 µm, which is in line with the experiments. Note however that the current PIC size is huge (2-3 cm) compared to commercially used chips (1 order of magnitude smaller). When using the flip-chip technique to integrate the much smaller chips in dedicated packages, warpage becomes less critical.

## 6. Conclusion

The "Mono" model for adiabatic taper design has been shown to be effective for a SiN-to-polymer transition. We have demonstrated that the heterogenous stack of CMOS and polymer materials can be obtained by using a face-up (lithography) and face-down (flip-chip bonding-approach, paving the way for face-up embedding and flip-chip integration of PICs. Both integration methods display similar results with adiabatic coupling efficiencies close to 1 dB and even lower, for both TE and TM polarizations near $\lambda = 1310$ nm.

Currently, the lithography approach provides more reproducible samples than flip-chip bonding, as the latter is more susceptible to warpage. Further optimization will be necessary to address this issue for large-scale production. Additionally, integrating tapered ORDL waveguides could further enhance coupling efficiency by fully leveraging the phase-matching condition with the tapered SiN, potentially allowing for shorter taper lengths.

This opens the possibility to use polymer material for CPO applications, based on the ORDL as optical communication channel. For both chip-to-chip coupling and chip-to-fiber coupling, a sub-2 dB loss can be obtained. By leveraging high-efficiency coupling between the PIC and ORDL, an ORDL pitch of less than 50 µm, and the fan-out towards the fiber array, a fiber-coupled package-to-package interconnect can be achieved.

**Funding**

European Union's Horizon Europe Research and Innovation Program under Agreement 101070560 (PUNCH).

**Acknowledgement**

Part of this work was supported by imec's industry-affiliation R&D Program on Optical I/O and by the European Union's Horizon Europe Research and Innovation Program under Agreement 101070560 (PUNCH). The authors would like to thank and acknowledge the CMST cleanroom support staff for fabrication support.

**Disclosures**

The authors declare no conflicts of interest.